%% file: snc.tex
\documentclass[11pt]{article}

\pdfoutput=1

\usepackage[margin=20mm]{geometry}
\input{header}

\numberwithin{equation}{section}
\numberwithin{figure}{section}

\usepackage{amssymb,amsmath}
\usepackage{graphicx}
\usepackage{rotating}
\usepackage{microtype}

\usepackage{caption}
\providecommand{\DeclareCaptionType}[1]{\relax}
\DeclareCaptionType{copyrightbox}

\usepackage{url}
\urldef{\mails}\path|{eric,asm,akobel,msagralo}@mpi-inf.mpg.de|

\newtheorem{lemma}{Lemma}

\newcommand{\bdesc}{$\textsc{Bdc}$~}

\begin{document}

\title{Arrangement Computation for Planar Algebraic Curves}

\author{Eric Berberich%
  \thanks{Max-Planck-Institut f\"ur Informatik, Saarbr\"ucken, Germany, \mails}
  \and Pavel Emeliyanenko$^*$
  \and Alexander Kobel$^*$
  \and Michael Sagraloff$^*$
}

\maketitle


\begin{abstract}
\input abstract
\end{abstract}


\section{Introduction}
\label{sec:intro}
\input introduction

\section{Curve Analysis}
\label{sec:curveanalysis}

\subsection{The Algorithm} 
\label{ssec:algorithm} 
\input algorithm

\subsection{Numerical Solver with Certificate}
\label{ssec:numerical}
\input numerical

\begin{figure*}[t]
  \centering
  \includegraphics[width=\linewidth]{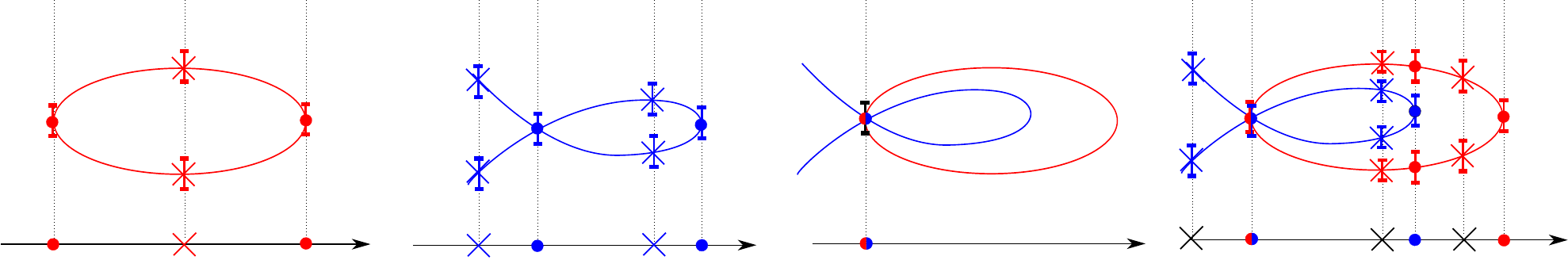}
\caption{The two figures on the left show the topology analyses for the curves $C=V(f)$ and $D=V(g)$. The figure in the middle shows the intersection of the two curves. For the curve pair analysis, critical event lines (at dots) are sorted and non-critical event lines (at crosses) in between are inserted. Finally, for each event line $x=\alpha$, the roots of $f(\alpha,y)$ and $g(\alpha,y)$ are sorted. The latter task is done by further refining corresponding isolating intervals (blue or red intervals) and using the combinatorial information from the curve analyses and the computation of the intersection points.\protect\\[-2em]\strut}
\label{fig:arrangement}
\end{figure*}

\subsection{Discussion}
\label{ssec:remarks}
\input fastliftremarks

\section{Arrangement Computation}
\label{sec:arrangement}
\input arrangements

\section{Implementation and Experiments}
\label{sec:experiments}

\input experiments


\section*{Acknowledgments}
Without Michael Kerber's careful implementation of the bivariate kernel in \textsc{Cgal}, this work would not have been realizable in a reasonable time. We would like to use the opportunity to thank Michael for his excellent work. 

\newpage

{
\bibliographystyle{abbrv}
\bibliography{bib,bib2,num}
}

\newpage
\begin{appendix}

\section{Further Experiments of Curve Analyses}
\label{asec:ca}
\input app_exp_ca

\ignore{
\newpage
\section{Further Experiments\\of Arrangements}
\label{asec:arr}
\input app_exp_arr
}

\newpage
\section{Description of Special Curves}
\label{asec:instances}
\input instances

\end{appendix}

\end{document}

%% file: header.tex
\usepackage{xspace}
\usepackage{paralist}
\usepackage{amsbsy,amssymb,amsmath,amsfonts}
\usepackage{fancybox}
\usepackage{color}
\usepackage{graphicx}
\usepackage{wrapfig}
\usepackage{url}
\usepackage{subfig}
\usepackage{verbatim}

\usepackage{array,tabularx,multirow}
\newcolumntype{L}{>{\raggedright\arraybackslash}X}
\newcolumntype{C}{>{\centering\arraybackslash}X}
\newcolumntype{R}{>{\raggedleft\arraybackslash}X}

\usepackage{algorithm}
\usepackage{algorithmic}

\usepackage{multibib}
\newcites{app}{Additional References}

\DeclareGraphicsExtensions{.png}
\DeclareGraphicsExtensions{.jpg}
\newcommand{\ignore}[1]{}

\newcommand{\Z}{\mathbb{Z}}
\newcommand{\RR}{\mathbb{R}}
\newcommand{\C}{\mathbb{C}}

\newcommand{\cgal}{\textsc{Cgal}\xspace}

\newcommand{\gmp}{\textsc{Gmp}\xspace}
\newcommand{\ntl}{\textsc{Ntl}\xspace}

\newcommand{\bs}{\textsc{Bisolve}\xspace}
\newcommand{\rs}{\textsc{Rs}\xspace}
\newcommand{\isolate}{\textsc{Isolate}\xspace}
\newcommand{\lgp}{\textsc{Lgp}\xspace}

\newcommand{\etal}{et~al.\xspace}

\newcommand{\fastlift}{\textsc{FastLift}\xspace}
\newcommand{\lift}{\textsc{Lift}\xspace}
\newcommand{\bisolve}{\textsc{Bisolve}\xspace}

\newcommand{\res}{\operatorname{res}}

\makeatletter
\def\fakemarginpar{%
  \ifhmode
    \@bsphack
    \@floatpenalty -\@Mii
  \else
    \@floatpenalty-\@Miii
  \fi
  \ifinner
    \@parmoderr
    \@floatpenalty\z@
  \else
    \@next\@currbox\@freelist{}{}%
    \@next\@marbox\@freelist{\global\count\@marbox\m@ne}%
       {\@floatpenalty\z@
        \@fltovf\def\@currbox{\@tempboxa}\def\@marbox{\@tempboxa}}%
  \fi
  \@ifnextchar [\@xmpar\@ympar}
\makeatother
\def\marrow{\fakemarginpar%
  [\textcolor{red}{\makebox[0pt][r]{$\longrightarrow$}}]%
  {\textcolor{red}{\makebox[0pt][l]{$\longleftarrow$}}}}

\def\eric#1{\textcolor{red}{\textsc{Eric says: }{\marrow\sf #1}}}

%% file: abstract.tex
We present a new \emph{certified} and \emph{complete} algorithm to compute 
arrangements of 
real planar algebraic curves. Our algorithm provides a 
geometric-topological analysis of the decomposition of the plane induced by a 
finite number of algebraic curves in terms of a cylindrical algebraic 
decomposition of the plane. Compared to previous approaches, we improve in 
two main aspects: 
Firstly, we significantly reduce the amount of exact operations, 
that is, our algorithms only uses resultant and $\gcd$ as purely symbolic 
operations. Secondly, we introduce a new hybrid method in the 
lifting step of our algorithm which combines the usage of a certified numerical complex root solver and information 
derived from the resultant computation. Additionally, we never consider any 
coordinate transformation and the output is also given with respect to the 
initial coordinate system.

We implemented our algorithm as a prototypical package of the C++-library 
\cgal. Our implementation exploits graphics hardware to expedite the 
resultant and $\gcd$ computation. We also compared our implementation with the current 
reference implementation, that is, \cgal's curve analysis and 
arrangement for algebraic curves. For various  
series of challenging instances, our experiments show that the new 
implementation outperforms the existing one.

%% file: introduction.tex
Computing the topology of a planar algebraic curve 
\begin{align}
C=V(f)=\{(x,y)\in\mathbb{R}^2:f(x,y)=0\}\label{def:curve}
\end{align}
can be considered as one of the fundamental problems in real algebraic geometry with numerous applications in computational geometry, computer graphics and computer aided geometric design. Typically, the topology of $C$ is given in terms of a planar graph $\mathcal{G}_{C}$ isotopic to $C$. For a geometric-topological analysis, we further require the vertices of $\mathcal{G}_{C}$ to be located on $C$. In this paper, we study the general problem of computing an arrangement of a given set of algebraic curves, that is, the decomposition of the plane into cells of dimensions $0$, $1$ and $2$ induced by the given curves. The proposed algorithm is \emph{certified} and \emph{complete}, and the overall arrangement computation is exclusively carried out in the initial coordinate system. Efficiency of our approach is shown by implementing our algorithm and comparing it to the current reference implementation.

There exist a number of certified and complete approaches to determine the topology of an algebraic curve; we refer the reader to~\cite{cheng.lazard.ea:on,eigenwilligkw07,gn-efficient,kerber-phd,LuisPhD2010} for recent work and further references. At present, only the method from~\cite{eigenwilligkw07} has been extended to arrangement computations of arbitrary algebraic curves~\cite{eigenwilligk08}. Common to all existing approaches is that, in a first step, they use eliminations techniques (e.g., resultants) to project the $x$-critical points (i.e., points $p\in C$ with $f_y(p)=0$) of the curve into one dimension. In a second step, the fiber at each of these projected points is computed. In general, this lifting step has turned out to be the most time-consuming part because it amounts to determining the real roots of a non-square univariate polynomial $f(\alpha,y)\in\mathbb{R}[y]$ with algebraic coefficients. The high computational cost for computing the roots of $f(\alpha,y)$ is mainly due to a more comprehensive algebraic machinery such as subresultants (in~\cite{eigenwilligk08,eigenwilligkw07,gn-efficient}), Gr\"obner basis or rational univariate representation (in~\cite{cheng.lazard.ea:on}) in order to obtain additional information on the number of distinct real (or complex) roots of $f(\alpha,y)$ or the multiplicity of the multiple roots of $f(\alpha,y)$. In addition, all except the method from~\cite{cheng.lazard.ea:on} consider a shearing of the curve which guarantees that the sheared curve has no two $x$-critical points sharing the same $x$-coordinate which in turn simplifies the lifting step but for the price of giving up sparseness of the initial input; an approach, which usually results in larger bitsizes of the coefficients and considerably increased running times. In a final connection step, arcs adjacent to the same $x$-critical points are identified.

The high-level description of the algorithm presented in this paper is almost identical to that of the existing methods, and, similar as in~\cite{eigenwilligk08,eigenwilligkw07}, we reduce the arrangement computation to the geometric-topological analysis of a single curve and of a pair of curves.
However, we improve in the following two main aspects: Firstly, we considerably reduce the amount of purely symbolic computations, that is, we only use resultant and $\gcd$ computation. The main reason for this approach is that we can outsource both computations to graphics hardware~\cite{emel-hpcs-11,emel-pasco-10,emel-ica3pp-10}, removing a bottleneck of previous methods which was due to the high amount of symbolic operations. Secondly, for curve analysis, we use a result from Teissier~\cite{Gwozdziewicz00formulaefor,Teissier} to obtain additional information for the number of distinct complex roots of $f(\alpha,y)$ along a critical fiber (actually, an upper bound which most likely matches the exact number). We combine this information with a new certified complex root solver~\cite{Kobel11} to isolate the roots of $f(\alpha,y)$. The latter symbolic-numeric step applies as an efficient filter denoted \fastlift which fails only for very special instances. In order to achieve completeness of our overall method (i.e., for cases where \fastlift fails), we modify the method from~\cite{bes-bisolve-2011} for solving bivariate polynomial systems in order to isolate the roots of $f(\alpha,y)$.

We implemented our algorithm as a development branch of  
\cgal's\footnote{Computational Geometry Algorithms Library,
\url{www.cgal.org}; see also \url{http://exacus.mpi-inf.mpg.de/cgi-bin/xalci.cgi} for an online demo on arrangement computation.} bivariate algebraic kernel (\textsc{Ak\_2} for short) which is based on the algorithms from~\cite{eigenwilligk08,eigenwilligkw07} for topology and arrangement computation. Intensive benchmarks~\cite{eigenwilligkw07,LuisPhD2010} have shown that \textsc{Ak\_2} can be considered as the current reference implementation. For fair comparison, we run an \textsc{Ak\_2} with GPU-enabled 
resultants and $\gcd$s against our implementation on numerous challenging benchmark instances.
Our experiments show that the new approach
outperforms \textsc{Ak\_2} for all instances. More 
precisely, our method is, on average, twice as fast for easy instances such as non-singular curves in generic position. 
For hard instances, 
we typically improve by large factors between $5$ and $120$ which is mainly due to the new symbolic-numeric filter \fastlift, the exclusive usage of resultant and $\gcd$ as only symbolic operations and the abstinence of shearing.
In summary, the presented approach demonstrates the strength of symbolic-numeric techniques. It further proves that, in order to achieve efficiency in an actual implementation, it is of great importance to reduce the amount of symbolic operations and to consider approximate operations whenever this is possible.

Finally, we are confident that our new approach will have some positive impact in the following respects: Existing subdivision 
methods~\cite{mourrain-sub,bcgy-complete,mp-subdivision} show excellent 
behavior for non-singular input. For singular input, they can be made 
certifying and complete when considering worst case separation bounds, however, this approach has not shown effective in practice so far. 
Another advantage of subdivision methods compared to elimination approaches 
is that they are local and do not need (global) algebraic operations. In our method, we considerably reduced the amount of such algebraic operations and outsourced the remaining computations to graphics card. Hence, it seems reasonable that combining our algorithm with a subdivision 
approach eventually leads to a certified \emph{and complete} method which shows excellent ``local'' behavior as well. 
We further see numerous applications of our method, in particular, when computing arrangements of surfaces. The actual implementation~\cite{bks-efficient} for surface triangulation is crucially based on planar arrangement computations of singular curves. Thus, we are confident to considerably improve its efficiency based on the new algorithm for planar arrangement computation.

%% file: algorithm.tex
The input of our algorithm is a planar algebraic curve
$C$ as defined in (\ref{def:curve}), where $f$ is a \emph{square-free}, bivariate polynomial $f\in\Z[x,y]$ with
integer coefficients. If $f$ is considered as polynomial in $y$ with
coefficients $f_i(x)\in\Z[x]$, its coefficients typically share a trivial content $h:=\gcd(f_0,f_1,\ldots)$.
A non-trivial content $h\notin\Z$ defines vertical lines at the real roots of $h$. Our algorithm handles this situation by dividing out $h$ first and finally merging the vertical lines defined by $h=0$ and the curve analysis of the curve $C':=V(f/h)$. Hence, throughout the following considerations, we can assume that $h$ is trivial, thus $C$ contains no vertical line.

The algorithm returns a planar graph $\mathcal{G}_C$
that is isotopic\footnote{$\mathcal{G}_C$ is isotopic to $C$ if there
exists a continuous mapping $\phi:[0,1]\times C\mapsto \RR^2$ such that
$\phi(0,C)=C$, $\phi(1,C)=\mathcal{G}_C$ and $\phi(t_0,.):C\mapsto
\phi(t_0,C)$ constitutes a homeomorphism for each $t_0\in[0,1]$.} to $C$,
where all vertices $V$ of $\mathcal{G}_C$ are located on $C$. The proposed
algorithm follows a classical cylindrical algebraic decomposition approach.
We start with a high-level description of the three-step algorithm.

In the first step, the \textbf{\emph{projection phase}}, we project all
\emph{$x$-critical points} $(\alpha,\beta)\in C$ (i.e.,
$f(\alpha,\beta)=f_y(\alpha,\beta)=0$) onto the $x$-axis by means of a
resultant computation and root isolation for the elimination polynomial.
The set of $x$-critical points comprises exactly the points where $C$ has a
vertical tangent or is singular. It is well known (e.g., see~\cite[Theorem 2.2.10]{kerber-phd} for a short proof) that, for any two
consecutive $x$-critical values $\alpha$ and $\alpha'$, $C$ is
\emph{delineable} over $I=(\alpha,\alpha')$, that is, $C|_{I\times\RR}$
decomposes into a certain number $m_I$ of disjoint function graphs
$C_{I,1},\ldots,C_{I,m_I}$. In the \textbf{\emph{lifting phase}}, we first isolate the roots of the (square-free)
\emph{intermediate polynomial} $f(q_I,y)\in\mathbb{Q}[y]$, where $q_I$
constitutes an arbitrary chosen but fixed rational value in $I$. This computation yields the number $m_I$ ($=$ number of real roots of $f(q_I,y)$) of arcs above $I$ and corresponding representatives
$(q_I,y_{I,i})\in C_{I,i}$ on each arc. We further compute all points on $C$ that are
located above an $x$-critical value $\alpha$, that is, we determine the
real roots $y_{\alpha,1},\ldots,y_{\alpha,m_{\alpha}}$ of each (non
square-free) \emph{fiber polynomial} $f(\alpha,y)\in\RR[y]$. From the
latter two computations, we obtain the vertex set $V$ of
$\mathcal{G}_C$ as the union of all points $(q_I,y_{I,i})$ and
$(\alpha,y_{\alpha,i})$. In the final \textbf{\emph{connection phase}},
which concludes the topology analysis, we determine which of the above
vertices are connected via an arc of $C$. For each connected pair
$(v_1,v_2)\in V$, we insert a line segment connecting $v_1$ and $v_2$. It
is then straightforward to prove that $\mathcal{G}_C$ is isotopic
to $C$; see~\cite[Theorem 6.4.4]{kerber-phd} for a proof. We remark that we never consider any kind of coordinate
transformation, even in case where $C$ contains two or more $x$-critical
points sharing the same $x$-coordinate.\\

\begin{figure*}[t]
  \centering
  \includegraphics[width=\linewidth]{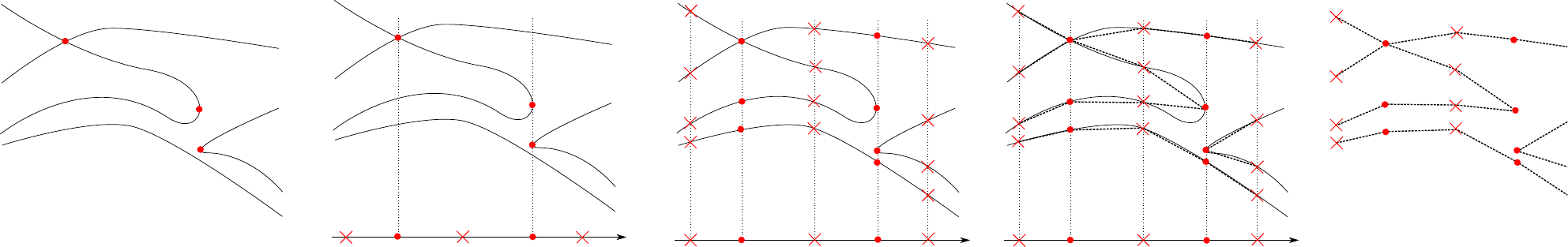}
\caption{\label{fig:cad}The figure on the left shows a curve $C$ with two
$x$-extremal points and one singular point (red dots). In the
\emph{projection phase}, these points are projected onto the $x$-axis and rational points separating the $x$-critical values are
inserted (red crosses). In the \emph{lifting phase}, the fibers at the
critical values (red dots) and at the rational points in between (red
crosses) are computed. In the \emph{connection phase}, each pair of lifted
points connected by an arc of $C$ is determined and a corresponding line
segment is inserted. The right figure shows the final graph that is
isotopic to $C$.\protect\\[-2em]\strut}
\end{figure*}

We next describe the three phases in detail:

\subsubsection{Projection Phase} 
\label{sssec:projection}

In the projection step, we follow well-known techniques from elimination
theory, that is, we compute the resultant $R(x):=\res(f,f_y,y) \in\Z[x]$
and a square-free factorization of $R$. More precisely, we determine
square-free and pairwise coprime factors $r_{i}\in\Z[x]$,
$i=1,\ldots,\deg(R)$, such that $R(x)=\prod_{i=1}^{\deg
(R)}\left(r_{i}(x)\right)^{i}$. We remark that, for some
$i\in\{1,\ldots,\deg(R)\}$, $r_{i}(x)\equiv 1$. Yun's
algorithm~\cite[Alg.~14.21]{gathen} constructs such a square-free
factorization by essentially computing greatest common divisors of $R$ and
its higher derivatives in an iterative way. Next, we isolate the real roots
$\alpha_{i,j}$, $j=1,\ldots,\ell_{i}$, of the polynomials $r_{i}$ which in
turn are $i$-fold roots of $R$. More precisely, we compute disjoint
intervals $I(\alpha_{i,j})\subset\RR$ with rational endpoints such that
$I(\alpha_{i,j})$ contains $\alpha_{i,j}$ but no other root of $r_{i}$, and
the union of all $I(\alpha_{i,j})$, $j=1,\ldots,\ell_{i}$, contains all
real roots of $r_{i}$. For the real root isolation, we consider the
Descartes method~\cite{vca,RS} as a suited algorithm. By further refining
the isolating intervals, we can achieve that all intervals
$I(\alpha_{i,j})$ are pairwise disjoint and, thus, also constitute
isolating intervals for the real roots of $R$. Then, for each pair $\alpha$
and $\alpha'$ of consecutive roots\ignore{\footnote{We artificially consider
$-\infty$ and $+\infty$ as the first and the last ``root''.}} of $R$
defining an open interval $I = (\alpha,\alpha')$, we choose a separating
rational value $q_I$ in between the corresponding isolating intervals.

\subsubsection{Lifting Phase}
\label{sssec:lifting}

Isolating the roots of the intermediate polynomials $f(q_I,y)$ is rather
straightforward: since $f(q_I,y)$ is a square-free polynomial with rational
coefficients, the Descartes method directly applies. Determining the
roots of $f(\alpha,y)\in\mathbb{R}[y]$ at an $x$-critical value $\alpha$ is
more complicated because $f(\alpha,y)$ has multiple roots and, in general,
irrational coefficients. We propose to run the following approach: We first
consider a method denoted $\fastlift$ which works as a filter for the fiber
computation. We will see in the experiments enlisted in
Section~\ref{sec:experiments} that $\fastlift$ applies to all fibers for the majority of
all input curves and only fails for a small number of fibers for some very special instances. In case of success, the fiber at $\alpha$ is returned. If $\fastlift$ fails, we use a second
method denoted $\lift$ which serves as a "backup" for $\fastlift$. In
comparison to $\fastlift$, $\lift$ is a complete method which
applies to any input curve and any corresponding $x$-critical value,
however, for the price of being less efficient. Nevertheless, our
experiments show that even the exclusive use of $\lift$ significantly improves upon
existing approaches.\\

\noindent\emph{\lift{} --- a complete method for fiber computation}:
\lift is based on our recent studies on solving a bivariate
polynomial system. In~\cite{bes-bisolve-2011}, we introduced a highly efficient method,
denoted \bisolve, to isolate the real solutions of a system of two
bivariate polynomials $f,g\in\mathbb{Q}[x,y]$. Its output consists of a set
of disjoint boxes $B_1,\ldots,B_m\subset\RR^2$ such that each box $B_i$
contains exactly one real solution $\xi:=(x_0,y_0)$ of $f(x,y)=g(x,y)=0$,
and the union of all $B_i$ covers all solutions. Furthermore, for each
solution $\xi$, \bisolve provides square-free polynomials
$p,q\in\Z[x]$ with $p(x_0)=q(y_0)=0$ and corresponding isolating (and
refineable) intervals $I(x_0)$ and $I(y_0)$ for $x_0$ and $y_0$,
respectively. Comparing $\xi$ with another point $\xi_1=(x_1,y_1)\in\RR^2$
given by a similar representation is rather straightforward. Namely, let
$\tilde{p},\tilde{q}\in\Z[x]$ be corresponding defining square-free
polynomials and $I(x_1)$ and $I(y_1)$ isolating intervals for $x_1$ and
$y_1$, respectively, then we can compare the $x$- and $y$-coordinates of
$\xi$ and $\tilde{\xi}$ via $\gcd$-computation of the defining univariate
polynomials and sign evaluation at the endpoints of the isolating intervals
(see~\cite[Algorithm 10.44]{bpr-algorithms} for more details).\\ 

In order to compute the fiber at an $x$-critical value $\alpha$ of $C$, we
proceed as follows: We first use \bisolve to determine all
solutions $p_i=(\alpha,\beta_i)$, $i=1,\ldots,l$, of the system $f=f_y=0$
with $x$-coordinate $\alpha$. Then, for each of these points, we compute 
\begin{align*}
k_i:=\min\{k:f_{y^k}(\alpha,\beta_i)=\frac{\partial^k f}{\partial
y^k}(\alpha,\beta_i)\neq 0\}\ge 2.
\end{align*}
The latter computation is done by iteratively calling \bisolve for
$f_y=f_{y^2}=0$, $f_{y^2}=f_{y^3}=0$, etc., and sorting the solutions along
the vertical line $x=\alpha$. We eventually obtain disjoint intervals
$I_1,\ldots,I_l$ and corresponding multiplicities $k_1,\ldots,k_l$ such
that $\beta_j$ is a $k_j$-fold root of $f(\alpha,y)$ which is contained in
$I_j$. The intervals $I_j$ already separate the roots $\beta_j$ from any
other multiple root of $f(\alpha,y)$, however, $I_j$ might still contain
ordinary roots of $f(\alpha,y)$. Hence, we further refine
each $I_j$ until we can guarantee via interval arithmetic that
$\frac{\partial^{k_{j}} f}{\partial y^{k_j}}(\alpha,y)$ does not vanish at $I_j$.
If this condition is fulfilled, $I_j$ cannot contain any root of
$f(\alpha,y)$ except $\beta_j$ due to the mean value theorem, thus, $I_j$
is an isolating interval.

After refining all intervals $I_j$, it remains to isolate the ordinary
roots of $f(\alpha,y)$. For this purpose, we use the so-called
\emph{Bitstream Descartes}  isolator~\cite{ek+-descartes} (\bdesc for
short) which can be considered as a variant of the Descartes method working
on polynomials with interval coefficients. This method can be used to get
arbitrary good approximations of the real roots of a polynomial with
``bitstream'' coefficients, that is, coefficients that can be approximated
to arbitrary precision. \bdesc starts from an interval guaranteed to
contain all real roots of a polynomial and proceeds with interval
subdivisions giving rise to a \emph{subdivision tree}. Accordingly, the
approximation precision for the coefficients is increased in each step of
the algorithm. Each leaf of the tree is associated with an interval $I$ and
stores a lower bound $l(I)$ and an upper bound $u(I)$ on the number of
real roots within this interval based on Descartes' Rule of Signs. Hence, $u(I)=0$ implies that $I$ contains no root and thus can be discarded. If $l(I)=u(I)=1$, then $I$ is an \emph{isolating interval} for a simple root. Intervals with $u(I)>1$ are further subdivided. We remark that,
after a number of iterations, \bdesc isolates all simple roots of a bitstream polynomial, and intervals not containing any root are eventually discarded. For a multiple root $\xi$, \bdesc constructs intervals $I$ which approximate $\xi$ to an arbitrary good precision but never certifies such an interval $I$ to be isolating.

In our situation, we have already isolated the multiple roots of $f(\alpha,y)$ by the intervals $I_j$. It remains to isolate the simple roots of $f(\alpha,y)$. Therefor, we consider the following modification of \bdesc: We discard an interval $I$ if one of following three cases applies:
\begin{inparaenum}[i)] \item $u(I)=0$, or \item $I$ is completely
contained in one of the intervals $I_j$, or \item $I$ contains an interval $I_j$
and $u(I)\le k_j$. \end{inparaenum}
Namely, in each of these situations, $I$ cannot contain
an ordinary root of $f(\alpha,y)$. An interval $I$ is stored as isolating
for an ordinary root of $f(\alpha,y)$ if $l(I)=u(I)=1$ and $I$ intersects
no interval $I_j$. All intervals which do not fulfill one of the above
conditions are further subdivided. In a last step, we sort the intervals $I_j$ (isolating the multiple roots) and the isolating intervals 
for the ordinary roots along the vertical line.\\

We remark that \bisolve applied in \lift reuses 
the resultant obtained in the projection phase of the algorithm.
Furthermore, it is a local approach in the sense that 
it suits very well for computing the fiber only at some specific 
$x$-critical value $\alpha$. Hence, when considering only a small number 
of $x$-critical fibers, the running time of \lift is considerably 
lower than its application to all fibers.\\

\noindent\emph{\fastlift{} --- a fast method for fiber computation}: $\fastlift$ is
a hybrid method to isolate all complex roots and, thus, also the real roots
of $f(\alpha,y)$, where $\alpha$ is an $x$-critical value of $C$. It
combines a numerical solver to compute arbitrary good approximations (i.e.,
complex discs in $\mathbb{C}$) of the roots, an exact certification step to
certify the existence of roots within the computed discs and the following
result due to Teissier~\cite{Gwozdziewicz00formulaefor,Teissier}:
\begin{lemma}[Teissier]
For an $x$-critical point $p=(\alpha,\beta)$ of $C=V(f)$, it holds that
\begin{align}
\operatorname{mult}(f(\alpha,y),\beta)=\operatorname{Int}(f,f_y,
p)-\operatorname{Int}(f_x,f_y,p)+1,\label{Teissier}
\end{align}
where $\operatorname{mult}(f(\alpha,y),\beta)$ denotes the multiplicity of
$\beta$ as root of $f(\alpha,y)\in\RR[y]$, $\operatorname{Int}(f,f_y,p)$
the intersection multiplicity\footnote{The intersection multiplicity of
two curves $f=0$ and $g=0$ at a point $p$ is defined as the dimension of
the localization of $\C[x,y]/(f,g)$ at $p$, considered as a $\C$-vector
space.} of the curves implicitly defined by $f=0$ and $f_y=0$ at $p$, and
$\operatorname{Int}(f_x,f_y,p)$ the intersection multiplicity of $f_x=0$
and $f_y=0$ at $p$. 
\end{lemma}

\noindent{\textbf{Remark.}} In the case, where $f_x$ and $f_y$ share a
common non-trivial factor $h=\gcd(f_x,f_y)\in\Z[x,y]$, $h$ does not vanish
at any $x$-critical point $p$ of $C$. Namely, $h(p)=0$ would imply that
$\operatorname{Int}(f_x,f_y,p)=\infty$ and, thus,
$\operatorname{Int}(f,f_y,p)=\infty$ as well, a contradiction to our
assumption on $f$ to be square-free. Hence, we have
$\operatorname{Int}(f_x,f_y,p)=\operatorname{Int}(f_x^*,f_y^*,p)$ with
$f_x^*:=f_x/h$ and $f_y^*:=f_y/h$ and, thus, the following formula (which
is equivalent to (\ref{Teissier}) for trivial $h$) applies: 
\begin{align}
\operatorname{mult}(f(\alpha,y),\beta)=\operatorname{Int}(f,f_y,
p)-\operatorname{Int}(f_x^*,f_y^*,p)+1,\label{Teissier2}
\end{align}

We come to the description of $\fastlift$. In the first step, we determine an upper bound $m_{\alpha}^*$
for the actual number $m_{\alpha}$ of distinct complex roots of
$f(\alpha,y)$ which most likely matches $m_{\alpha}$. We distinguish the cases $\deg f(\alpha,y)\neq\deg_y f$ and $\deg f(\alpha,y)=\deg_y f$. In the first case, $C$ has a vertical asymptote at $\alpha$. Then, we define $m_{\alpha}^*:=\deg f(\alpha,y)$ which is obviously an upper bound for $m_{\alpha}$. If $C$ is in generic position, $f(\alpha,y)$ has only ordinary roots and, thus, $m_{\alpha}^*=m_{\alpha}$.

We now consider the case $\deg f(\alpha,y)=\deg_y f$. Then, due to the
formula (\ref{Teissier2}), we have
\begin{align}\nonumber
m_{\alpha}&=\#\{\text{distinct complex roots of }f(\alpha,y)\}\\ \nonumber
&=\deg_y f-\deg\gcd(f(\alpha,y),f_y(\alpha,y))\\ \nonumber
&=\sum_{\beta\in\C:f(\alpha,\beta)=0}(\operatorname{mult}(f(\alpha,y),
\beta)-1)\\ \nonumber
&=\deg f_y - \sum_{\substack{\beta\in\C:\\
    \makebox[2em][c]{\scriptsize $(\alpha,\beta)$ is $x$-critical}}}
\left(\operatorname{Int}(f,f_y,(\alpha,\beta))-\operatorname{
Int}(f_x^*,f_y^*,(\alpha,\beta))\right)\\
&=\deg_y f-\operatorname{mult}(R,\alpha)+ \sum_{\substack{\beta\in\C:\\
    \makebox[2.5em][c]{\scriptsize $(\alpha,\beta)$ is $x$-critical}}}
\operatorname{Int}(f_x^*,f_y^*,(\alpha,\beta))\label{eq1}\\
&\le\deg_y
f-\operatorname{mult}(R,\alpha)+\sum_{\beta\in\C}\operatorname{Int}(f_x^*,
f_y^*,(\alpha,\beta)) \label{ineq1}\\
&=\deg_y
f-\operatorname{mult}(R,\alpha)+\operatorname{mult}(Q,\alpha)=:m_{\alpha}^*\label{eq2}
\end{align}
where $R(x)=\operatorname{res}(f,f_y,y)$ and
$Q(x):=\res(f_x^*,f_y^*,y)$. The equality (\ref{eq1}) is due
to the fact that $f$ has no vertical asymptote at $\alpha$ and, thus, the
multiplicity $\operatorname{mult}(R,\alpha)$ equals the sum
$\sum_{\beta\in\C}\operatorname{Int}((f,f_y,(\alpha,\beta))$ of the
intersection multiplicities of $f$ and $f_y$ in the fiber at $\alpha$. (\ref{eq2}) follows by an analogous argument for the
intersection multiplicities of $f_x^*$ and $f_y^*$ along the vertical line
at $\alpha$. From the square-free factorization of $R$, we already know
$\operatorname{mult}(R,\alpha)$, and $\operatorname{mult}(Q,\alpha)$ can be
determined by computing $Q$, its square-free factorization and checking
whether $\alpha$ is a root of one of the factors. 
If the curve $C$ is in generic position\footnote{The reader may notice that generic position is used in a different context here. It is required that all intersection points of $f_x^*$ and $f_y^*$ above $\alpha$ are located on the curve $C$. We remark that this requirement is usually fulfilled even if there are two $x$-critical values sharing the same $x$-coordinate.}, the inequality (\ref{ineq1})
becomes an equality because then $f_x^*$ and $f_y^*$ do not intersect in
any point above $\alpha$ which is not located on $C$. Thus, in this case, we have $m_{\alpha}=m^*_{\alpha}$.\\

\begin{figure*}[t]
  \centering
  \includegraphics[width=0.9\linewidth]{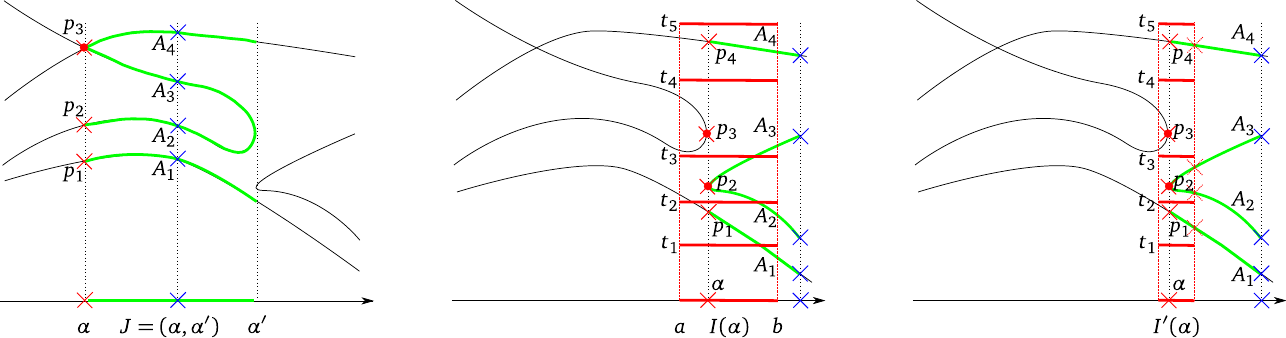}
\caption{\label{fig:connection}The left figure shows the generic case, where
exactly one $x$-critical point ($p_3$) above $\alpha$ exists.
  The bottom-up method connects $A_1$ to $p_1$ and $A_2$ to $p_2$; the
remaining arcs have to pass $p_3$. In the second figure, the fiber at
$\alpha$ contains two critical points $p_2$ and $p_3.$
  The red horizontal line segments pass through arbitrary chosen points
$(\alpha,t_i)$ separating $p_{i-1}$ and $p_i$.
  The initial isolating interval $I(\alpha) = (a,b)$ for $\alpha$ is not
sufficient to determine the connections for all arcs since $A_1, A_2, A_3$
intersect the segments $I \times \{ t_i \}.$ On the right, the refined
isolating interval $I'(\alpha)$ induces boxes $I'(\alpha) \times (t_i,
t_{i+1})$ small enough such that no arc crosses the horizontal boundaries.
By examination of the $y$-coordinates of the intersections between the arcs
and the fiber over the right-hand boundary of $I'(\alpha)$ (red crosses),
we can match arcs and critical points.\protect\\[-2em]\strut}
\end{figure*}

In the second step of \fastlift, we aim to isolate all (complex) roots
of $f(\alpha,y)$. The above computation of an upper bound $m_{\alpha}^*$ motivates the following ansatz: In order to
determine the roots of $f(\alpha,y)$, we combine a numerical complex solver
for $f(\alpha,y)$ and an exact certification step. More precisely, we use
the numerical solver to determine disjoint discs $D_1,\ldots,D_{m}$ in the
complex space and an exact certification step to certify the existence of a
certain number $m_i\ge 1$ of roots (counted with multiplicity) of
$f(\alpha,y)$ within each $D_i$; see Section~\ref{ssec:numerical} for
details. Increasing the working precision and the number of iterations for
the numerical solver eventually leads to arbitrary well refined discs $D_i$
-- but without a guarantee that these discs are actually isolating! Since
$m_{\alpha}^*$ constitutes an upper bound on the number of distinct complex
roots of $f(\alpha,y)$, we must have $m\le m_{\alpha}\le m_{\alpha}^*$ at
any time. Hence, if the number of discs $m$ equals the upper bound
$m_{\alpha}^*$, we know for sure that all complex roots of $\deg
f(\alpha,y)$ are isolated. Then, the isolating discs $D_1,\ldots,D_m$ are further refined until, for all
$i=1,\ldots,m$, 
\begin{align}
D_i\cap\mathbb{R}= \emptyset\text{ or }\bar{D}_i\cap D_j=\emptyset\text{
for all }j\neq i,\label{condition1} 
\end{align}
where $\bar{D}_i:=\{\bar{z}:z\in D_i\}$ denotes the complex conjugate of
$D_i$. The latter condition guarantees that each disc $D_i$ which
intersects the real axis actually isolates a real root of $f(\alpha,y)$. In
addition, for each real root isolated by some $D_i$, we further
obtain its multiplicity $m_i$ as a root of $f(\alpha,y)$.

If $m \neq m_{\alpha}^*$, either one of the roots of
$f(\alpha,y)$ is still not isolated or it holds $m_{\alpha}<m_{\alpha}^*$.
In the case that we do not succeed, that is, we still have $m<m_{\alpha}^*$ 
after a number of iterations (in the implementation, this number is set empirically) with increasing precision, we stop
\fastlift and proceed with \lift instead. In Section~\ref{ssec:remarks},
we discuss the few failure cases.

We remark that the usage of Teissier's formula is not entirely new when computing the topology of algebraic curves. In~\cite{cheng.lazard.ea:on,LuisPhD2010}, the formula (\ref{Teissier}) was used in its simplified form to compute $\operatorname*{mult}(\beta,f(\alpha,y))$ for a non-singular point $p=(\alpha,\beta)$ (i.e., $\operatorname*{Int}(f_x,f_y,p)=0$). In contrast, we use the formula in its general form and sum up the information along the entire fiber which eventually leads to the upper bound $m_{\alpha}^*$ on the number of distinct complex roots of $f(\alpha,y)$.

\subsubsection{Connection Phase} 
\label{sssec:connection}

Let us consider a fixed $x$-critical value $\alpha$, the corresponding
isolating interval $I(\alpha)=(a,b)$ computed in the projection phase and
the points $p_i:=(\alpha,y_{\alpha,i})\in C$, $i=1,\ldots,m_{\alpha}$,
located on $C$ above $\alpha$. Furthermore, let $I=(\alpha,\alpha')$ be the
interval connecting $\alpha$ with the nearest $x$-critical value to the
right of $\alpha$ (or $+\infty$ if none exists) and $A_j$, $j=1,\ldots,m_I$, the $j$-th arc of $C$ above
$I$ with respect to vertical ordering. To its left, $A_j$ is either
connected to $(\alpha,\pm \infty)$ (in case of a vertical asymptote) or to
one of the points $p_i$. In order to determine the point to which an arc
$A_j$ is connected, we consider the following two distinct cases:
\begin{itemize}
\item The \emph{generic case}, in which there exists exactly one real
$x$-critical point $p_{i_0}$ above $\alpha$ and $\deg f(\alpha,y)=\deg_y
f$. The latter condition implies that $C$ has no vertical asymptote at
$\alpha$. Then, the points $p_1,\ldots,p_{i_0-1}$ must be connected with
$A_1,\ldots,A_{i_0-1}$ in button-up fashion, respectively, since, for each of these points,
there exists a single arc of $C$ passing this point. The same argument
shows that $p_{i_0+1},\ldots,p_{m_{\alpha}}$ must be connected to
$A_{m_I-m_{\alpha}+i_0+1},\ldots,A_{m_I}$ in top-down fashion, respectively. Finally, the remaining arcs
in between must all be connected to the $x$-critical point $p_{i_0}$.

\item The \emph{non-generic case}: Each arc $A_j$ is represented by a point
$a_j:=(q_I,y_{I,j})\in C$, where $y_{I,j}$ denotes the $j$-th real root of
$f(q_I,y)$ and $q_I$ an arbitrary but fixed rational value in $I$. We choose
arbitrary rational values $t_1,\ldots,t_{m_{\alpha}+1}$ with
$t_1<y_{\alpha,1}<t_2<\ldots<y_{\alpha,m_{\alpha}}<t_{m_{\alpha}+1}$. Then,
the points $\tilde{p}_i:=(\alpha,t_i)$ separate the $p_i$'s from each
other. Computing such $\tilde{p}_i$ is easy since we have isolating
intervals with rational endpoints for each of the roots $y_{\alpha,i}$ of
$f(\alpha,y)$. In a second step, we use interval arithmetic to obtain
intervals $\mathfrak{B}f(I(\alpha)\times t_i)$ with $f(I(\alpha)\times
t_i)\subset \mathfrak{B}f(I(\alpha)\times t_i)$. As long as there exists an
$i$ with $0\in \mathfrak{B}f(I(\alpha)\times t_i)$, we refine $I(\alpha)$.
Since none of the $\tilde{p}_i$ is located on $C$, we eventually obtain a
sufficiently refined interval $I(\alpha)$ with $0\notin
\mathfrak{B}f(I(\alpha)\times t_i)$ for all $i$. It follows that none of
the arcs $A_j$ intersects any line segment $I(\alpha)\times t_i$. Hence,
above $I(\alpha)$, each $A_j$ stays within the the rectangle bounded by the
two segments $I(\alpha)\times t_{i_0}$ and $I(\alpha)\times t_{i_0+1}$ and
is thus connected to $p_{i_0}$. In order to determine $i_0$, we compute the
$j$-th real root $\gamma_j$ of $f(b,y)\in\mathbb{Q}[y]$ and the largest
$i_0$ such that $\gamma_j>t_{i_0}$. In the special case where
$\gamma_j<t_i$ or $\gamma_j>t_i$ for all $i$, it follows that $A_j$ is
connected to $(\alpha,-\infty)$ or $(\alpha,+\infty)$, respectively.
\end{itemize} 

For the arcs located to the left of $\alpha$, we proceed in exactly the
same manner. This concludes the connection phase and, thus, the description
of our algorithm.

%% file: numerical.tex
\newcommand{\abs}[1]{\ensuremath{\left|#1\right|}}%
\newcommand{\PN}{\ensuremath{\operatorname{\text{\textit{PN}}}}}%
\newcommand{\RN}{\ensuremath{\operatorname{\text{\textit{RN}}}}}%
In \fastlift, we deploy a certified numerical solver for a fiber polynomial to find regions certified to contain its complex roots.
Bini and Fiorentino presented a highly efficient solution to this problem in their \textsc{MPSolve} package \cite{BF00-design}. 
We adapt their approach in a way suited to handle the case where the coefficients are not known a priori, but rather in an intermediate representation which can be refined to any arbitrary finite precision.
The description given in this section is high-level.
For the details of an efficient implementation, we refer the reader to \cite{Kobel11}.
In the following considerations, let $g(z) := f(\alpha, z) = \sum_{i=0}^n g_i z^i \in \RR[z]$ be the fiber polynomial at an $x$-critical value $\alpha$ and $V(g) = \{ \zeta_i \}$, $i = 1, \dots, n,$ its complex roots. Thus, $g(z) = g_n \allowbreak \prod_{i=1}^n \allowbreak (z - \zeta_i)$.

The main algorithm used in the numerical solver is the Aberth-Ehrlich iteration for simultaneous root finding.
Starting from arbitrary distinct root guesses $(z_i)_{i=1,\dots,n}$, it is given by the component-wise iteration rule $z'_i = z_i$ if $g(z_i) = 0$, and
\begin{equation*}
  z'_i = z_i - \frac{g(z_i) / g'(z_i)}{1 - g(z_i) / g'(z_i) \cdot \sum_{j \ne i} \frac{1}{z_i - z_j}}
\end{equation*}
otherwise. As soon as the approximation vector $(z_i)_i$ lies in a sufficiently small neighborhood of some permutation of the actual roots $(\zeta_i)_i$ of $g$, this iteration converges with cubic order \cite{Tilli98}.
In practice, Aberth's method shows excellent performance even if started from an arbitrary configuration far away from the solutions.

A straightforward implementation of \cite{BF00-design} expects the coefficients $g_i$ of $g$ to be known up to some relative precision $p$, that is, the input is a polynomial $\tilde{g} = \sum \tilde{g_i} x^i$ whose floating point coefficients satisfy $\abs{\tilde{g_i}-g_i} \le 2^{-p} \abs{g_i}$.
In particular, this requirement implies that we have to decide in advance whether a coefficient vanishes.
In general, though, the critical $x$-coordinate $\alpha$ of the fiber polynomial, and thus the coefficients of $g$, are not rational.
Thus, it translates to expensive symbolic $\gcd$ computations of $R$ and the coefficients of $f$ as a univariate polynomial in $\Z[y][x]$.

Instead, we work on a \emph{Bitstream interval representation} \cite{Eigenwillig08,Kobel11} $[g]^\mu$ of $g$.
Its coefficients are interval approximations of the coefficients of $g$, where we require the width $|g_i^+ - g_i^-|$ of each coefficient $[g]^\mu_i = [g_i^-, g_i^+]$ to be $\le \mu$ for a certain \emph{absolute} precision $\mu = 2^{-p}$.
In this sense $[g]^\mu$ represents the set $\{ \tilde{g} : \tilde{g_i} \in [g]^\mu_i \}$ of polynomials in a \emph{$\mu$-polynomial neighborhood} of $g$; in particular, $g$ itself is contained in $[g]^\mu$.
Naturally, for the interval boundaries, we consider dyadic floating point numbers (\emph{bigfloats}).
Note that we can easily compute arbitrarily good Bitstream representations of $f(\alpha, z)$ by approximating $\alpha$ to an arbitrary small error, for example using the quadratic interval refinement technique \cite{Abbott06}.

%

Starting with some precision (say, $\mu = 2^{-53}$) and a vector of initial approximations, we perform Aberth's iteration on some representant $\tilde{g} \in [g]^\mu$.
The natural choice is the \emph{median polynomial} with $\tilde{g_i} = (g_i^- + g_i^+)/2$, but we take the liberty to select other candidates in case of numerical singularities in Aberth's rule (most notably, if $\tilde{g}'(z_i) = 0$
in some iteration).

After a finite number of iterations (depending on the degree of $g$), we interrupt the iteration and check whether the current approximation state already captures the structure of $V(g)$.
We use the following result by Neumaier and Rump \cite{Rump03}, founded in the conceptually similar Wei\-er\-stra\ss-Du\-rand-Ker\-ner simultaneous root iteration:
\begin{lemma}[Neumaier]
  Let $g(z) = g_n \prod_{i=1}^n (z - \zeta_i) \in \C[z]$, $g_n \ne 0$.
  Let $z_i \in \C$ for $i = 1, \dots, n$ be pairwise distinct root approximations.
  Then, all roots of $g$ belong to the union $\mathcal{D}$ of the discs
  \begin{gather*}
    D_i := D(z_i - r_i, |r_i|),\\
    \text{where } r_i := \frac{n}{2} \cdot \frac{\omega_i}{g_n} \text{ and } \omega_i := \frac{g(z_i)}{\prod\nolimits_{j \ne i} (z_i - z_j)}.
  \end{gather*}
  Moreover, every connected component $C$ of $\mathcal{D}$ consisting of $m$ discs contains \emph{exactly} $m$ zeros of $g$, counted with multiplicity.
\end{lemma}
The above lemma applied to $[g]^\mu$ using conservative interval arithmetic yields a superset $\mathcal{C} = \{ C_1, \dots, C_m \}$ of regions and corresponding multiplicities $\lambda_1, \dots, \lambda_m$ such that, for each $C_k \in \mathcal{C}$, all polynomials $\tilde{g} \in [g]^\mu$ (and, in particular, $g$) have exactly $\lambda_k$ roots in $C_k$ counted with multiplicities.
Furthermore, once the quality of the approximations $(z_i)_i$ and $[g]^\mu$ is sufficiently high, $\mathcal{C}$ converges to $V(g)$.

In \fastlift, where we aim to isolate the roots of $g:=f(\alpha,y)$, we check whether $m=m_{\alpha}^{*}$. If the latter equality holds, Teissier's lemma guarantees that the regions $C_k \in \mathcal{C}$ are isolating for the roots of $g$, and we stop.
Otherwise, we repeat Aberth's iteration after checking whether $0 \in [g]^\mu(z_i)$.
Informally, if this holds the quality of the root guess is not distinguishable from any (possibly better) guess within the current interval approximation of $g$, and we double the precision ($\mu' = \mu^2$) for the next stage.


Aberth's iteration lacks a proof for convergence in the general case and, thus, cannot be considered complete.
However, we feel this is a merely theoretical issue: to the best of our knowledge, only artificially constructed, highly degenerate configurations of initial approximations render the algorithm to fail.
Regardless of this assumption, the regions $C_k \in \mathcal{C}$ are certified to comprise the roots of $g$ at any stage of the algorithm by Neumaier's lemma and the rigorous use of interval arithmetic.
Furthermore, since we use the numerical method as a filter only, the completeness of the overall approach is not harmed.

%

%% file: fastliftremarks.tex
We conclude our description of the curve analysis algorithm with the following discussion on our method \fastlift in the lifting step.
\fastlift is a certified method, that is, in case of success, it returns the mathematical correct result. However, the reader may notice that \fastlift does not apply in all cases, a reason why we additionally consider the complete but less efficient backup method \lift for some of the $x$-critical fibers.
The failure of \fastlift is either due to a \emph{very special geometric situation} along a certain fiber or due to the \emph{behavior of the numerical solver}. Special geometric situations are:
\begin{enumerate}
\item[(Geo1)] $C$ has a vertical tangent at $x=\alpha$ and $f(\alpha,y)$ is not square-free, or
\item[(Geo2)] there exists an intersection point of $f_x^{*}=\frac{f_{x}}{\gcd(f_{x},f_{y})}$ and $f_y^{*}=\frac{f_{y}}{\gcd(f_{x},f_{y})}$ above an $x$-critical value $\alpha$ of $f$ which is not located on $C$.
\end{enumerate}

In case that none of the above special geometric situations is given (or removed by applying a shear; see below), 
the success of \fastlift is guaranteed if the following conditions on the numerical solver are fulfilled:

\begin{enumerate}
\item[(Num1)] The numerical solver is run for sufficiently many iterations, and
\item[(Num2)] the approximations returned by the numerical sol\-ver converge against the roots of $f(\alpha,y)$ when increasing precision and number of iterations.
\end{enumerate}

As already mentioned in Section~\ref{ssec:numerical}, we consider (Num2) as an exclusively theoretical problem. We further remark that, alternatively, we can use an exact and complete complex bitstream solver~\cite{s-bitstream-mcs11} to compute arbitrary good approximations of the fiber polynomial $f(\alpha,y)$, an approach which can be considered as an extension of \bdesc to complex roots. However, for efficiency reasons, we decided to integrate a numerical method into our implementation instead. (Num1) is a practical problem and, in our implementation, we just empirically set the maximal number of iterations and the maximal precision as parameters depending on the degree and the bitsize of the polynomials. We noticed that the success of \fastlift actually depends on these parameter values and consider it an interesting research question how they should be related to the given input to achieve optimal running times.

It is worth mentioning that a complete and exact topology computation can be fully based on \fastlift if we allow shearing (i.e., a coordinate transformation $x\mapsto x+sy$ for a $s\in\mathbb{Q}$). Namely, for all but except a finite number $N$ of shearing factors,\footnote{In~\cite[Corollary 3.2.10]{kerber-phd}, it is shown that there exist at most $n^4+n$ shearing factors such that the sheared curve has covertical $x$-critical points. In analogous manner, one can compute an upper bound for the number of shearing factors, where the sheared curve is not in (Geo1) or (Geo2).} (Geo1) and (Geo2) do not apply to the sheared curve. Hence, when considering $N+1$ pairwise distinct shearing factors one by one in circular order and increasing the number of iterations and the precision in the numerical solver for each factor, \fastlift eventually succeeds for all fibers.

In practice, as observed in our experiments presented in Section~\ref{sec:experiments}, 
the failure conditions for \fastlift are almost negligible, as the method only fails for a few critical fibers on very special instances. For the remaining
fibers and for all fibers of the majority of instances, \fastlift is successful and extremely fast.

%% file: arrangements.tex
\cgal{}'s recent implementation for computing arrangements
of planar algebraic curves reduces all required geometric constructions
(as intersections) and predicates (as comparisons of points
and $x$-monotone curves) to the geo\-me\-tric-to\-po\-lo\-gi\-cal
analysis of a single curve~\cite{eigenwilligkw07} and pairs of 
curves~\cite{eigenwilligk08}; see also~\cite{bhk-ak2-2011} and 
\cgal{}'s documentation~\cite{cgal:wfzh-a2-10b}.

Beyond the improved curve analysis proposed in Section~\ref{sec:curveanalysis}, 
we aim to avoid subresultant sequences in general 
when analysing pairs of curves (see illustration in Figure~\ref{fig:arrangement}), 
which is straightforward given the analyses of each single curve and 
the common intersection points of the two curves  computed by \bisolve:

\ignore{
In Section~\ref{ssec:algorithm}, we proposed a new algorithm for the analysis of a single curve that replaces the existing approach
and thus avoids the costly computations of subresultant sequences
for singular input curves or curves in non-generic position. 
The following consideration shows that the analysis
of a pair of curves is rather straightforward given the analyses of each single curve 
and the common intersection points of the two curves as computed by \bisolve.
Again, using \bisolve removes the need for subresultants in a curve-pair
analysis that we illustrate in Figure~\ref{fig:arrangement}.
}

Let $C=V(f)$ and $D=V(g)$ be two planar algebraic curves implicitly defined by the square-free polynomials $f$, $g\in\mathbb{Z}[x,y]$. The curve analysis for $C$ provides a set of \emph{critical event lines} 
$x=\alpha$, where $f(\alpha,y)$ is non square-free. Each $\alpha$ is 
represented as the root of a square-free polynomial $r_i$, with $r_i$ a factor 
of $R_C:=\operatorname(f,f_y,y)$, together with an isolating interval 
$I(\alpha)$. In addition, we have isolating intervals for the roots of 
$f(\alpha,y)$. A corresponding result also holds for the curve $D$ with $R_D:=\operatorname*{res}(g,g_y,y)$. For 
the common intersection points of $C$ and $D$, a similar representation is known. 
That is, we have critical event lines $x=\alpha'$, where $\alpha'$ is a root of a square-free 
factor of $R_{CD}:=\operatorname{res}(f,g,y)$ and, thus, $f(\alpha,y)$ and 
$g(\alpha,y)$ share at least one common root (or the their leading coefficients both vanish for $x=\alpha$). In addition, isolating 
intervals for each of these roots are computed. The curve-pair analysis now 
essentially follows from merging this information. More precisely, we first 
compute merged \emph{critical event lines} (via sorting the roots of $R_C$, $R_D$ and $R_{CD}$) and, then, pad merged \emph{non-critical event lines} at rational values $q_I$ in between. The 
intersections of $C$ and $D$ with a non-critical event line at $x=q_I$ are easily 
computed by isolating the roots of $f(q_I,y)$ and $g(q_I,y)$ and further 
refining the isolating intervals until all isolating intervals are pairwise disjoint. 
For a critical event line $x=\alpha$, we refine the already computed isolating 
intervals for $f(\alpha,y)$ and $g(\alpha,y)$ until the number of pairs of 
overlapping intervals matches the number $m$ of intersection points of $C$ and $D$ 
above $\alpha$. This number is obtained from the output of \bisolve
applied to $f$ and $g$, restricted to $x=\alpha$. The information on how to connect the lifted points is provided by the curve analyses for $C$ and $D$.

We remark that, in the previous approach by Eigenwillig and Kerber~\cite{eigenwilligk08}, $m$ is determined via efficient filter methods, too, while in general, a
subresultant computation is needed if the filters fail. This is, for instance, the case when two covertical intersections of $C$ and $D$ occur.

%% file: experiments.tex
\textbf{\textit{Setup.}} We have implemented our algorithms 
as a branch of the bivariate
algebraic kernel released with version~3.7 in October 2010, and replaced
the curve and curve-pair analyses therein with our new methods based on
\fastlift, \lift and \bisolve. As throughout \cgal, we follow the
\emph{generic programming paradigm}, which allows us to choose among
various number types and methods to isolated the real roots of integral
univariate polynomials. For our setup, we rely on the number types provided
by \gmp~5.0.1\footnote{\gmp:~\url{http://gmplib.org}} and the highly
efficient univariate solver based on the Descartes method contained in \rs by
Fabrice~Rouillier~\cite{RS},\footnote{\rs:~\url{http://www.loria.fr/equipes/vegas/rs}}
which is also the basis for \isolate in Maple~13.

All experiments have been conducted on 2.8~GHz $8$-Core Intel Xeon W3530 with 8~MB
of L2~cache on a Linux platform. For the GPU-part of the algorithm, we
have used the GeForce GTX580 graphics card (Fermi Core).
All used data sets are available for download.\footnote{
\url{http://www.mpi-inf.mpg.de/departments/d1/projects/Geometry/DataSetsSNC-2011.zip}}\\

\textit{\textbf{Symbolic Speedups.}}
Our algorithm exclusively relies on two symbolic operations, that is, resultant and $\gcd$ computation. We outsource 
\emph{both} computations to the graphics hardware to reduce the overhead of 
symbolic arithmetic which typically constitutes the main bottleneck in previous approaches. Besides a quick introduction given next, 
we refer the interested reader 
to~\cite{emel-hpcs-11,emel-pasco-10,emel-ica3pp-10} for more details.

Shortly, both approaches are based on the ``di\-vide-con\-quer-com\-bine'' principle
used in the modular algorithms by Brown \cite{brown-71} and Collins~\cite{collins-71}. 
This principle allows to distribute the computation over a large
number of processor cores of the graphics card. At highest level, the
modular approach can be formulated as follows. \begin{inparaenum}[1.]\item Apply
modular and/or evaluation homomorphisms to reduce the problem to a large
set of subproblems over a simple domain. \item Solve the subproblems
individually in a finite field. \item Recover the result with 
polynomial interpolation (in case evaluation homomorphism has been applied) and
Chinese remaindering.\end{inparaenum}

Altogether, the GPU realization is quite straightforward and does not
deserve much attention within the context of this work. It is only worth
noting that our implementation of univariate $\gcd$s on the graphics card is
comparable in speed with the one from \ntl{}\footnote{\ntl,
\url{http://www.shoup.net/ntl}}~5.5 running on the host machine. Our intuition
for this is because, in contrast to bivariate resultants, computing a $\gcd$
of moderate degree univariate polynomials does not provide a
sufficient amount of parallelism, and \ntl's implementation is nearly
optimal. Moreover, the time for initial modular reduction of polynomials,
still performed on the CPU, can become noticeably large, thereby
neglecting the efficiency of the GPU algorithm.
Yet, we find it very promising to perform the modular reduction 
on the GPU which should further speed-up our algorithm.\\

\textsl{\textbf{Contestants.}}
We compare the new implementation with \cgal's bivariate algebraic kernel
(see \cite{bhk-ak2-2011} and \cite{cgal:bht-ak-10}) that has shown 
excellent performance in exhaustive experiments 
over existing approaches, namely 
\textsc{cad2d}\footnote{\url{http://www.usna.edu/Users/cs/qepcad/B/QEPCAD.html}} and 
\textsc{Isotop}~\cite{gn-efficient} which is based on \rs.
Both contestants were, except for few example instances, less efficient than 
\cgal's implementation, so that we omit further tests with them. 
Two further reasons can be given: Firstly, we enhanced \cgal{}'s kernel with 
GPU-supported resultants and $\gcd$s, which makes it 
more competitive to existing software, but also to our new software, in case of non-singular curves, though
slowdowns are still expected for singular curves or curves in non-generic
position due to the need of subresultants sequences performed on the CPU.
Secondly, the contestants based on \rs require as subtask \rs to solve
the bivariate polynomial system $f = f_y = 0$ in the curve-analysis.
In \cite{bes-bisolve-2011}, we learned that the GPU-supported 
\bisolve was at least competitive to the current version of \rs and even showed in most cases
an excellent speed gain over \rs. However, \rs is currently getting a 
very budding polish based on Rational Univariate Representations and 
modular arithmetic. Its theory will be presented at 
EuroCG 2011~\cite{blpr-rs3-eurocg-11}. We are looking 
forward to compare our algorithms with a preliminary version of the new \rs
quite soon. Moreover, we are confident that the authors will support 
the computation of arrangements of algebraic curves in a similar setup.
Though tackling the same problem, both realizations can be seen as 
orthogonal if not complementary.

\subsection{Curve Analysis}
\label{ssec:curveanalysis}

We first present the experiments comparing the analyses of single algebraic
curves for different families of curves: 
\begin{inparaenum}
\item[(R)] random curves of various degree and bit-lengths of their coefficients,
\item[(I)] curves interpolated through points on a grid,
\item[(S)] curves in the two-dimensional parameter space of a sphere,
\item[(T)] curves that were constructed by multiplying a curve $f(x,y)$ with $f(x,y+1)$, 
such that each fiber has more than one critical point.
\item[(P)] projections of intersections of algebraic surfaces in 3D and, finally,
\item[(X)] ``special'' curves of degrees up to 42 with many singularities or high-curvature points. We already considered these special curves in~\cite{bes-bisolve-2011} and describe them in more detail
in Table~\ref{tbl:app:special-desc} in Appendix~\ref{asec:ca}.
\end{inparaenum}
The non-random and non-special curves are taken from~\cite[4.3]{kerber-phd}.
For the curve topology analysis, we consider four different
setups: 
\begin{inparaenum}[(a)]
 \item \bisolve is, strictly speaking, not comparable with 
the actual curve analysis as it only computes the solutions
of the system $f = f_y = 0$. 
Still, it is interesting to see that our new hybrid method even outperforms
this algorithm that only solves a subproblem of the curve-analysis.
\item \textsc{LiftAna} exclusively uses \lift for the fiber computations.
\item \textsc{FastAna} combines \fastlift and \lift in the fiber computations. More precisely, it uses \fastlift first, and if it fails for a certain fiber, \lift is considered for this fiber instead,
\item \textsc{Ak\_2} is the bivariate algebraic kernel shipped with \cgal~3.7,
but with GPU-supported resultants and $\gcd$s.
\end{inparaenum}
\textsc{FastAna} is our default setting, and its running time also includes the timing for the fiber computations where \fastlift fails and \lift is applied instead.

\begin{table}[t!]
\sffamily\scriptsize
\caption{Running times (in sec) for analyses of algebraic curves of various families; \textbf{timeout:} algorithm timed out ($>$~400~sec)}
\label{tbl:cana}
\centering
\begin{tabularx}{\linewidth}{ |>{\hsize=1.4\hsize}R| |>{\hsize=0.85\hsize}R| |>{\hsize=0.975\hsize}R|>{\hsize=0.975\hsize\bfseries}R| |>{\hsize=0.8\hsize}R| } 

\hline
\multicolumn{5}{|l|}{\multirow{2}{*}{(R)~sets of five random curves}}\\
\multicolumn{5}{|c|}{} \\
\hline
degree, bits & \normalfont\bisolve & \normalfont\textsc{LiftAna} & \normalfont\textsc{FastAna} & \normalfont\textsc{Ak\_2} \\
\hline
\multicolumn{5}{|l|}{dense curves}\\
\hline
9, \hphantom{00}10   &0.83   &0.73   &0.24   &0.66\\
9, 2048 &4.17   &9.00   &2.24   &3.43\\
15, \hphantom{00}10  &2.82   &3.47   &1.01   &2.21\\
15, 2048&20.48  &50.06  &13.31  &16.82\\
\hline
\multicolumn{5}{|l|}{sparse curves}\\
\hline
9, \hphantom{00}10   &0.25   &0.33   &0.11   &1.00\\
9, 2048 &0.25   &0.39   &0.15   &0.98\\
15, \hphantom{00}10  &1.50   &3.31   &0.57   &3.19\\
15, 2048&15.66  &32.15  &8.08   &24.32\\

\hline
\multicolumn{5}{|l|}{\multirow{2}{*}{(I)~curve interpolated through points on a grid}}\\
\multicolumn{5}{|c|}{} \\
\hline
degree & \normalfont\bisolve & \normalfont\textsc{LiftAna} & \normalfont\textsc{FastAna} & \normalfont\textsc{Ak\_2} \\
\hline
9 &4.51   &7.25   &2.50   &4.98\\
12 &25.25  &45.58  &14.66  &27.07\\

\hline
\multicolumn{5}{|l|}{\multirow{2}{*}{(S)~parametrized curve on a sphere with 16bit-coefficients}}\\
\multicolumn{5}{|c|}{} \\
\hline
degree & \normalfont\bisolve & \normalfont\textsc{LiftAna} & \normalfont\textsc{FastAna} & \normalfont\textsc{Ak\_2} \\
\hline
6 &6.52   &8.17   &2.49   &14.65\\
9 &51.92  &78.09  &24.14  &45.66\\

\hline
\multicolumn{5}{|l|}{\multirow{2}{*}{(T)~curves wih a vertically translated copy}}\\
\multicolumn{5}{|c|}{} \\
\hline
degree & \normalfont\bisolve & \normalfont\textsc{LiftAna} & \normalfont\textsc{FastAna} & \normalfont\textsc{Ak\_2} \\
\hline
6   &5.27   &3.15   &0.83   &12.89\\
9   &14.78  &14.91  &2.97   &159.22\\

\hline
\multicolumn{5}{|l|}{\multirow{2}{*}{(P)~projected~intersection~curve of
surfaces with 8bit-coefficient}}\\
\multicolumn{5}{|c|}{} \\
\hline
degree(s) & \normalfont\bisolve & \normalfont\textsc{LiftAna} & \normalfont\textsc{FastAna} & \normalfont\textsc{Ak\_2} \\
\hline
$2\cdot2$ &0.26   &0.23   &0.09   &0.22\\
$3\cdot3$   &2.87   &1.75   &0.52   &1.97\\
$4\cdot4$   &10.01  &11.93  &2.24   &38.91\\
$5\cdot5$  &83.00  &95.37  &13.76  & timeout\\

\hline
\multicolumn{5}{|l|}{\multirow{2}{*}{(X)~special~curves (see Table~\ref{tbl:app:special-desc}, Appendix \ref{asec:instances} for descriptions)}}\\
\multicolumn{5}{|c|}{} \\
\hline
name & \normalfont\bisolve & \normalfont\textsc{LiftAna} & \normalfont\textsc{FastAna} & \normalfont\textsc{Ak\_2} \\

\hline

curve\_issac    &2.89   &2.30   &0.44   &2.96\\
SA\_2\_4\_eps   &0.35   &2.18   &0.64   &73.00\\
grid\_deg\_10   &1.32   &2.52   &0.79   &1.54\\
L6\_circles     &4.52   &92.24  &2.10   &224.19\\
huge\_cusp      &8.50   &18.40  &5.50   &16.10\\
swinnerton      &19.61  &16.81  &7.12   &304.09\\
degree\_7\_surf &14.57  &48.94  &7.39   &timeout\\
spider          &57.09  &195.21 &26.26  &timeout\\
FTT\_5\_4\_4    &11.70  &44.27  &32.23  &timeout\\
challenge\_12b  &16.48  &52.71  &44.30  &timeout\\
mignotte\_xy    &260.61 &270.70 &136.54 &timeout\\

\hline
\end{tabularx}
\end{table}

Table~\ref{tbl:cana} lists the running times for single-curve analyses.
We only give the results for representative examples; 
complete tables are given in Appendix~\ref{asec:ca}.
It is easy to see that \fastlift is generally 
superior to the existing kernel, even though \cgal's implementation 
profits from GPU-accelerated symbolic arithmetic.
Moreover, while the speed-up for curves in generic position is already considerable 
(about half of the time), it becomes even more impressive for projected intersection curves
of surfaces and ``special'' curves with singularities.
The reason for this tremendous speed-up is that, for singular 
curves, \textsc{Ak\_2}'s performance drops significantly with the degree
of the curve when the time to compute subresultants on the CPU becomes the dominating
bottleneck of that approach. In addition, for curves in 
non-generic position, the efficiency of \textsc{Ak\_2} is affected because 
a coordinate transformation has to be considered in these cases.

Recall that the symbolic-numerical filter in \fastlift fails for very few instances, 
in which case \lift is locally used instead. 
The switch to the backup method is observable in timings; see
for instance, \emph{challenge\_12b}.
As a result, the difference of the running times between \lift and \fastlift are considerably less than for instances where the filter method succeeds for all fibers. 
In these cases, the numerical solver cannot isolate the roots
within a given number of iterations, 
or we indeed have $m < m^*_{\alpha}$; see~Section~\ref{ssec:remarks}.
Nevertheless, the running times are still very promising
and yet perform better than \textsc{Ak\_2} for non-generic input,
even though \lift's implementation is not yet mature enough,
and we anticipate a further performance improvement.

Similar as \textsc{Ak\_2} has improved on previous approaches when it was presented in 2008, our
new methods improve on \textsc{Ak\_2} now. That is, for random, interpolated 
and pa\-ra\-me\-tri\-zed curves, the speed gain is noticeable, while
for translated curves and projected intersections, we improve the more the higher the degrees. On special curves of large degree(!), we improve by a 
factors up to 100 and more.


\subsection{Arrangements}
\label{ssec:arrangements}

For arrangements of algebraic curves, we compare two implementations:
\begin{inparaenum}[(A)]
 \item \textsc{Ak\_2} is \cgal{}'s bivariate algebraic kernel shipped with
\cgal~3.7 but with GPU-supported resultants and gcds. 
 \item \textsc{FastKernel} is the same, but relies on \fastlift-filtered
analyses of single algebraic curves. For the curve pair analysis, \textsc{FastKernel} exploits \textsc{Ak\_2}'s functionality whenever subresultant computations are not needed (i.e., a unique transversal intersection of two curves along a critical event line). For more difficult situations (i.e., two covertical intersections or a tangential intersection), the curve pair analysis
uses \bisolve as explained in 
Section~\ref{sec:arrangement}.
\end{inparaenum}
Our testbed consists of sets of curves from different families:
\begin{inparaenum}
\item[(F)] random rational functions of various degree
\item[(C)] random circles
\item[(E)] random ellipses
\item[(R)] random curves of various degree and coefficient bit-length
\item[(P)] sets of projected intersection curves of algebraic surfaces
\ignore{
\item[(A)] planar arrangement used to analyze an algebraic surface
via projection (includes subresultant sequence of surface) following
the approach in~\cite{bks-efficient} where the efficiency of the projection
constitutes the bottleneck
}, and, finally,
\item[(X)] combinations of ``special'' curves.
\end{inparaenum}

\begin{wraptable}[20]{r}{0.53\linewidth}
\vspace{-7.5mm}
\sffamily\scriptsize
\caption{Running times (in sec) for computing arrangements of algebraic curves; \textbf{timeout:} algorithm timed out ($>$~4000~sec)}
\label{tbl:arr}
\centering
\begin{tabularx}{\linewidth}{ |>{\hsize=1.3\hsize}R| |>{\bfseries}R| |>{\hsize=0.7\hsize}R| } 

\hline
\multicolumn{3}{|l|}{\multirow{2}{*}{(P)~increasing number of projected surface intersections}}\\
\multicolumn{3}{|c|}{} \\
\hline
\#resultants & \normalfont\textsc{FastKernel} & \normalfont\textsc{Ak\_2} \\
\hline
 2   &0.21   &0.49\\
 3   &0.48   &0.93\\
 4   &1.03   &1.64\\
 5   &2.44   &3.92\\
 6   &5.14   &7.84\\
 7   &13.65  &21.70\\
 8   &22.69  &35.77\\
 9   &41.53  &67.00\\
10   &58.37  &91.84\\

\ignore{
\hline
\multicolumn{3}{|l|}{\multirow{2}{*}{(A)~arrangement of (sub)resultants for surface analysis}}\\
\multicolumn{3}{|c|}{} \\
\hline
instance, degree & \normalfont\textsc{FastKernel} & \normalfont\textsc{Ak\_2} \\
\hline
rand, 4 & TODO & & \\
interpolated, 4 & TODO & & \\
tangle\_cube, 555 & TODO & & \\
}

\hline
\multicolumn{3}{|l|}{\multirow{2}{*}{(X)~combinations of special curves (see Table \ref{tbl:app:special-desc}, Appendix \ref{asec:instances})}}\\
\multicolumn{3}{|c|}{} \\
\hline
\#curves & \normalfont\textsc{FastKernel} & \normalfont\textsc{Ak\_2} \\
\hline
 2   &9.2    &81.93\\
 3   &25.18  &148.46\\
 4   &248.87   &730.57\\ 
 5   &323.42 &836.43\\
 6   &689.39 &3030.27\\
 7   &757.94 &3313.27\\
 8   &1129.98    &timeout\\
 9   &1166.17    &timeout\\
10   &1201.34    &timeout\\
11   &2696.15    &timeout\\

\hline
\end{tabularx}
\end{wraptable}
We skip the tables for rational functions, circles, ellipses and
random curves because the performance of both
contestants are more or less equal: The \emph{linearly} many 
curve-analyses are simple and, for the \emph{quadratic} number of curve-pair
analyses, there are typically no multiple intersections along a fiber,
that is, \bisolve is not triggered. Thus, the execution paths of
both implementations are almost identical, but only as we enhanced 
\textsc{Ak\_2} with GPU-enabled resultants and $\gcd$s. In addition, we also do not expect the need of a shear for such curves, thus, the behavior is
anticipated. The picture changes for projected intersection curves of 
surfaces and combinations of special curves whose running times are 
reported in~Table~\ref{tbl:arr}. The \textsc{Ak\_2} requires for both sets 
expensive subresultants to analyze single curves and to compute covertical 
intersections, while \textsc{FastKernel}'s performance is crucially less affected in such situations.

%% file: app_exp_ca.tex
\begin{table}[H]
\sffamily\scriptsize
\caption{Running times (in sec) for analyses of random algebraic curves; \textbf{timeout:} algorithm timed out ($>$~400~sec)}
\label{tbl:app:random}
\centering
\begin{tabularx}{\linewidth}{ |>{\hsize=1.4\hsize}R| |>{\hsize=0.85\hsize}R| |>{\hsize=0.975\hsize}R|>{\hsize=0.975\hsize\bfseries}R| |>{\hsize=0.8\hsize}R| } 

\hline
\multicolumn{5}{|l|}{\multirow{2}{*}{(R)~sets of five random curves}}\\
\multicolumn{5}{|c|}{} \\
\hline
degree, bits & \normalfont\bisolve & \normalfont\textsc{LiftAna} & \normalfont\textsc{FastAna} & \normalfont\textsc{Ak\_2} \\
\hline

\multicolumn{5}{|l|}{dense curves}\\
\hline
6, \hphantom{00}10   &0.38   &0.39   &0.15   &0.36\\
6, \hphantom{0}128   &0.32   &0.49   &0.18   &0.32\\
6, \hphantom{0}512   &0.55   &0.89   &0.35   &0.55\\
6, 2048              &1.85   &3.50   &0.99   &1.71\\
\hline
9, \hphantom{00}10   &0.83   &0.73   &0.24   &0.66\\
9, \hphantom{0}128   &0.57   &0.90   &0.31   &0.55\\
9, \hphantom{0}512   &1.04   &1.91   &0.57   &0.96\\
9, 2048              &4.17   &9.00   &2.24   &3.43\\
\hline
12, \hphantom{00}10  &2.18   &2.21   &0.69   &1.77\\
12, \hphantom{0}128  &1.64   &2.87   &0.86   &1.46\\
12, \hphantom{0}512  &2.84   &6.09   &1.56   &2.55\\
12, 2048             &12.07  &30.26  &7.06   &9.96\\
\hline
15, \hphantom{00}10  &2.82   &3.47   &1.01   &2.21\\
15, \hphantom{0}128  &2.36   &4.55   &1.29   &2.06\\
15, \hphantom{0}512  &4.40   &9.84   &2.63   &3.77\\
15, 2048             &20.48  &50.06  &13.31  &16.82\\
\hline
\multicolumn{5}{|l|}{sparse curves}\\
\hline
6, \hphantom{00}10   &0.17   &0.21   &0.12   &0.22\\
6, \hphantom{0}128   &0.14   &0.19   &0.09   &0.22\\
6, \hphantom{0}512   &0.21   &0.36   &0.12   &0.39\\
6, 2048              &0.73   &1.13   &0.39   &1.06\\
\hline
9, \hphantom{00}10   &0.25   &0.33   &0.11   &1.00\\
9, \hphantom{0}128   &0.25   &0.39   &0.15   &0.98\\
9, \hphantom{0}512   &0.43   &0.65   &0.23   &1.43\\
9, 2048              &1.53   &2.51   &0.85   &4.31\\
\hline
12, \hphantom{00}10  &0.41   &0.83   &0.19   &1.67\\
12, \hphantom{0}128  &0.40   &1.05   &0.25   &1.60\\
12, \hphantom{0}512  &0.78   &2.16   &0.50   &2.53\\
12, 2048             &3.82   &10.54  &2.82   &9.94\\
\hline
15, \hphantom{00}10  &1.50   &3.31   &0.57   &3.19\\
15, \hphantom{0}128  &1.48   &3.98   &0.72   &2.99\\
15, \hphantom{0}512  &2.96   &7.10   &1.45   &5.58\\
15, 2048             &15.66  &32.15  &8.08   &24.32\\

\hline
\end{tabularx}
\end{table}

\newpage

\begin{table}[H]
\sffamily\scriptsize
\caption{Running times (in sec) for analyses of algebraic curves of various families; \textbf{timeout:} algorithm timed out ($>$~400~sec)}
\label{tbl:app:inter}
\centering
\begin{tabularx}{\linewidth}{ |>{\hsize=1.4\hsize}R| |>{\hsize=0.85\hsize}R| |>{\hsize=0.975\hsize}R|>{\hsize=0.975\hsize\bfseries}R| |>{\hsize=0.8\hsize}R| } 

\hline
\multicolumn{5}{|l|}{\multirow{2}{*}{(I)~curve interpolated through points on a grid}}\\
\multicolumn{5}{|c|}{} \\
\hline
degree & \normalfont\bisolve & \normalfont\textsc{LiftAna} & \normalfont\textsc{FastAna} & \normalfont\textsc{Ak\_2} \\
\hline

5  &0.41   &0.50   &0.21   &0.49\\
6  &0.72   &1.13   &0.41   &0.84\\
7  &1.42   &2.20   &0.78   &1.62\\
8  &2.59   &3.78   &1.32   &2.86\\
9  &4.51   &7.25   &2.50   &4.98\\
10 &6.62   &11.52  &3.62   &7.65\\
11 &12.38  &23.27  &7.25   &13.93\\
12 &25.25  &45.58  &14.66  &27.07\\
13 &48.97  &93.97  &27.90  &46.21\\
14 &101.96 &193.61 &59.14  &90.27\\
15 &211.95 &timeout&114.68 &166.68\\
16 &timeout&timeout&236.39 &314.61\\

\ignore{
\hline
\end{tabularx}
\end{table}

\begin{table}[H]
\sffamily\small
\caption{Running times (in sec) for analyses of parameterized curves on a sphere; \textbf{timeout:} algorithm timed out ($>$~400~sec)}
\label{tbl:app:param}
\centering
\begin{tabularx}{\linewidth}{ |>{\hsize=1.4\hsize}R| |>{\hsize=0.85\hsize}R| |>{\hsize=0.975\hsize}R|>{\hsize=0.975\hsize\bfseries}R| |>{\hsize=0.8\hsize}R| } 
}

\hline
\multicolumn{5}{|l|}{\multirow{2}{*}{(S)~parametrized curve on a sphere with 16bit-coefficients}}\\
\multicolumn{5}{|c|}{} \\
\hline
degree & \normalfont\bisolve & \normalfont\textsc{LiftAna} & \normalfont\textsc{FastAna} & \normalfont\textsc{Ak\_2} \\
\hline
 6  &6.52   &8.17   &2.49   &14.65\\
 7  &22.58  &27.39  &8.47   &16.83\\
 8  &37.94  &53.3   &16.25  &32.09\\
 9  &51.92  &78.09  &24.14  &45.66\\
10  &72.21  &110.81 &32.38  &64.31\\

\ignore{
\hline
\end{tabularx}
\end{table}

\addtocounter{table}{-1}
\begin{table}[H]
\sffamily\small
\caption{Running times (in sec) for analyses of curves with covertical events; \textbf{timeout:} algorithm timed out ($>$~400~sec)\protect\\(continued)\protect\\[0em]\strut}
\label{tbl:app:trans}
\centering
\begin{tabularx}{\linewidth}{ |>{\hsize=1.4\hsize}R| |>{\hsize=0.85\hsize}R| |>{\hsize=0.975\hsize}R|>{\hsize=0.975\hsize\bfseries}R| |>{\hsize=0.8\hsize}R| } 
}

\hline
\multicolumn{5}{|l|}{\multirow{2}{*}{(T)~curves with a vertically translated copy}}\\
\multicolumn{5}{|c|}{} \\
\hline
degree & \normalfont\bisolve & \normalfont\textsc{LiftAna} & \normalfont\textsc{FastAna} & \normalfont\textsc{Ak\_2} \\
\hline
 5  &3.95   &1.93   &0.6    &5.61\\
 6  &5.27   &3.15   &0.83   &12.89\\
 7  &7.31   &6.14   &1.28   &33.97\\
 8  &9.66   &7.99   &1.75   &73.59\\
 9  &14.78  &14.91  &2.97   &159.22\\
10  &17.03  &15.78  &3.36   &timeout\\

\hline
\multicolumn{5}{|l|}{\multirow{2}{*}{(P)~projected~intersection~curve of
surfaces with 8bit-coefficient}}\\
\multicolumn{5}{|c|}{} \\
\hline
degree(s) & \normalfont\bisolve & \normalfont\textsc{LiftAna} & \normalfont\textsc{FastAna} & \normalfont\textsc{Ak\_2} \\
\hline

$2\cdot2$    &0.26   &0.23   &0.09   &0.22\\
$2\cdot3$    &0.39   &0.40   &0.12   &0.32\\
$2\cdot4$    &1.03   &0.77   &0.25   &0.73\\
$2\cdot5$    &1.97   &1.53   &0.44   &1.93\\
$3\cdot3$    &2.87   &1.75   &0.52   &1.97\\
$3\cdot4$    &3.63   &3.02   &0.67   &5.21\\
$3\cdot5$    &8.71   &9.22   &1.90   &25.87\\
$4\cdot4$    &10.01  &11.93  &2.24   &38.91\\
$4\cdot5$    &21.45  &28.22  &4.46   &231.64\\
$5\cdot5$    &83.00  &95.37  &13.76  &timout\\

\ignore{
\hline
\end{tabularx}
\end{table}

\begin{table}[H]
\sffamily\small
\caption{Running times (in sec) for analyses of special algebraic curves; \textbf{timeout:} algorithm timed out ($>$~400~sec)}
\label{tbl:app:special}
\centering
\begin{tabularx}{\linewidth}{ |>{\hsize=1.4\hsize}R| |>{\hsize=0.85\hsize}R| |>{\hsize=0.975\hsize}R|>{\hsize=0.975\hsize\bfseries}R| |>{\hsize=0.8\hsize}R| } 
}

\hline
\multicolumn{5}{|l|}{\multirow{2}{*}{(X)~special~curves (see Table \ref{tbl:app:special-desc}, Appendix \ref{asec:instances} for desciptions)}}\\
\multicolumn{5}{|c|}{} \\
\hline
name & \normalfont\bisolve & \normalfont\textsc{LiftAna} & \normalfont\textsc{FastAna} & \normalfont\textsc{Ak\_2} \\

\hline


curve\_issac    &2.89   &2.30   &0.44   &2.96\\
L4\_circles     &1.56   &8.80   &0.52   &9.92\\
SA\_2\_4\_eps   &0.35   &2.18   &0.64   &73.00\\
grid\_deg\_10   &1.32   &2.52   &0.79   &1.54\\
ten\_circles    &6.77   &3.79   &0.82   &24.07\\
dfold\_10\_6    &4.45   &4.62   &1.58   &30.80\\
L6\_circles     &4.52   &92.24  &2.10   &224.19\\
cov\_sol\_20    &8.68   &12.16  &3.38   &65.05\\
curve24         &14.07  &15.61  &4.07   &50.73\\
huge\_cusp      &8.50   &18.40  &5.50   &16.10\\
swinnerton      &19.61  &16.81  &7.12   &304.09\\
degree\_7\_surf &14.57  &48.94  &7.39   &timeout\\
challenge\_12a  &7.47   &15.84  &14.41  &timeout\\
spider          &57.09  &195.21 &26.26  &timeout\\
FTT\_5\_4\_4    &11.70  &44.27  &32.23  &timeout\\
challenge\_12b  &16.48  &52.71  &44.30  &timeout\\
mignotte\_xy    &260.61 &270.70 &136.54 &timeout\\

\hline
\end{tabularx}
\end{table}

%% file: app_exp_arr.tex
\vspace{-6.5cm}

\begin{table}[h]
\sffamily\small
\caption{Running times (in sec) for computing arrangements of algebraic curves}
\label{tbl:app:arr}
\centering
\begin{tabularx}{0.9\linewidth}{ |>{\hsize=1.3\hsize}R| |>{\bfseries}R| |>{\hsize=0.7\hsize}R| } 

\hline
\multicolumn{3}{|l|}{\multirow{2}{*}{(F)~rational functions }}\\
\multicolumn{3}{|c|}{} \\
\hline
\#functions, degree & \normalfont\textsc{FastKernel} & \normalfont\textsc{Ak\_2} \\
\hline
20, \hphantom{0}2 &0.41   &0.38\\
20, \hphantom{0}3 &0.51   &0.49\\
20, \hphantom{0}4 &0.70   &0.62\\
20, \hphantom{0}5 &0.76   &0.77\\
\hline
20, \hphantom{0}6 &0.88   &0.82\\
30, \hphantom{0}6 &1.82   &1.72\\
40, \hphantom{0}6 &2.95   &2.89\\
50, \hphantom{0}6 &4.52   &4.30\\
60, \hphantom{0}6 &6.46   &6.30\\
70, \hphantom{0}6 &9.16   &8.61\\
80, \hphantom{0}6 &11.52  &10.75\\
90, \hphantom{0}6 &14.72  &14.04\\
\hline
20, \hphantom{0}7 &0.99   &0.93\\
20, \hphantom{0}8 &1.06   &1.00\\
20, \hphantom{0}9 &1.22   &1.25\\
20, 10            &1.38   &1.33\\
20, 11            &1.40   &1.40\\
20, 12            &1.49   &1.48\\
20, 13            &1.76   &1.74\\
20, 14            &1.66   &1.63\\
20, 15            &1.93   &1.92\\

\ignore{
\hline
\end{tabularx}
\end{table}

\addtocounter{table}{-1}
\begin{table}[H]
\sffamily\small
\caption{Running times (in sec) for computing arrangements of algebraic curves (continued)}
\label{tbl:app:arr}
\centering
\begin{tabularx}{0.9\linewidth}{ |>{\hsize=1.3\hsize}R| |>{\bfseries}R| |>{\hsize=0.7\hsize}R| } 
}

\hline
\multicolumn{3}{|l|}{\multirow{2}{*}{(C)~random circles}}\\
\multicolumn{3}{|c|}{} \\
\hline
\#circles & \normalfont\textsc{FastKernel} & \normalfont\textsc{Ak\_2} \\
\hline
 100 &    6.22  &   5.98\\
 200 &   24.45  &  23.15\\
 300 &   49.62  &  46.77\\
 400 &   78.54  &  73.15\\
 500 &  123.00  & 114.16\\

\hline
\multicolumn{3}{|l|}{\multirow{2}{*}{(E)~random ellipses}}\\
\multicolumn{3}{|c|}{} \\
\hline
\#ellipses & \normalfont\textsc{FastKernel} & \normalfont\textsc{Ak\_2} \\
\hline
10  &   0.16  &  0.16\\
20  &   0.43  &  0.42\\
30  &   0.95  &  0.92\\
40  &   1.67  &  1.60\\
50  &   2.36  &  2.18\\
60  &   3.25  &  2.91\\
70  &   4.07  &  3.77\\
80  &   5.15  &  4.89\\
90  &   6.30  &  5.73\\
200 &  27.05  & 25.04\\

\hline
\end{tabularx}
\end{table}

\pagebreak

\addtocounter{table}{-1}
\begin{table}[H]
\sffamily\small
\caption{Running times (in sec) for computing arrangements of algebraic curves
\hfill(continued)}
\label{tbl:app:arr}
\centering
\begin{tabularx}{0.9\linewidth}{ |>{\hsize=1.3\hsize}R| |>{\bfseries}R| |>{\hsize=0.7\hsize}R| } 

\hline
\multicolumn{3}{|l|}{\multirow{2}{*}{(R)~sets of 5 random curves}}\\
\multicolumn{3}{|c|}{} \\
\hline
degree, bits & \normalfont\textsc{FastKernel} & \normalfont\textsc{Ak\_2} \\
\hline

\multicolumn{3}{|l|}{dense curves}\\
\hline
6, \hphantom{00}10   &0.56   &0.61\\
6, \hphantom{0}128   &0.60    &0.68\\
6, \hphantom{0}512   &0.86   &0.99\\
6, 2048              &2.75   &3.16\\
\hline               
9, \hphantom{00}10   &2.18   &2.44\\
9, \hphantom{0}128   &2.42   &2.56\\
9, \hphantom{0}512   &3.43   &3.68\\
9, 2048              &11.50   &12.39\\
\hline               
12, \hphantom{00}10  &9.17   &9.98\\
12, \hphantom{0}128  &9.55   &9.91\\
12, \hphantom{0}512  &11.88  &12.69\\
12, 2048             &31.76  &34.26\\
\hline               
15, \hphantom{00}10  &28.15  &29.52\\
15, \hphantom{0}128  &29.28  &30.22\\
15, \hphantom{0}512  &34.04  &34.06\\
15, 2048             &80.89  &82.61\\

\hline
\multicolumn{3}{|l|}{sparse curves}\\
\hline
6, \hphantom{00}10   &0.28   &0.36\\
6, \hphantom{0}128   &0.30    &0.44\\
6, \hphantom{0}512   &0.44   &0.64\\
6, 2048              &1.40    &2.00\\
\hline               
9, \hphantom{00}10   &1.01   &2.04\\
9, \hphantom{0}128   &1.13   &1.89\\
9, \hphantom{0}512   &1.56   &2.97\\
9, 2048              &5.16   &9.66\\
\hline               
12, \hphantom{00}10  &2.83   &4.62\\
12, \hphantom{0}128  &3.20    &4.80 \\
12, \hphantom{0}512  &4.99   &7.20 \\
12, 2048             &23.19  &29.15\\
\hline               
15, \hphantom{00}10  &15.46  &18.42\\
15, \hphantom{0}128  &15.97  &19.36\\
15, \hphantom{0}512  &19.93  &23.86\\
15, 2048             &57.30   &70.64\\

\hline
\end{tabularx}
\end{table}

%% file: instances.tex
\begin{table}[H]
  \centering\sffamily\scriptsize
  \caption{Description of the special curves used in the experiments of Section~\ref{sec:experiments}. Sources
of curves are given where known.}
  \label{tbl:app:special-desc}
  \begin{minipage}[t]{\linewidth}
    \centering
    \begin{tabularx}{\linewidth}{|l|r|L|}
      \hline
      Instance & $y$-degree & Description\\
      \hline
      curve\_issac      &    15 & isolated points, high-curvature points \cite{LGP-09}\\
      L4\_circles       &    16 & 4 circles w.r.t.\ L4-norm;\newline clustered solutions\\
      SA\_2\_4\_eps     &    16 & singular points with high tangencies, displaced \cite{labs_10}\\
      grid\_deg\_10     &    10 & large coefficients;\newline curve in generic position\\
      ten\_circles      &    20 & set of 10 random circles multiplied together; rational solutions\\
      dfold\_10\_6      &    30 & many half-branches \cite{labs_10}\\
      L6\_circles       &    32 & 4 circles w.r.t.\ L6-norm;\newline clustered solutions\\
      cov\_sol\_20      &    20 & covertical solutions\\
      curve24           &    24 & curvature of degree 8 curve;\newline many singularities\\
      huge\_cusp        &     8 & large coefficients;\newline high-curvature points\\
      swinnerton        &    25 & covertical solutions in $x$ and $y$\\ 
      degree\_7\_surf   &    42 & silhouette of an algebraic surface; covertical solutions in $x$ and $y$\\ 
      challenge\_12a    &    30 & many candidate solutions to be checked \cite{labs_10}\\
      spider            &    12 & degenerate curve;\newline many clustered solutions\\
      FTT\_5\_4\_4      &    40 & many non-rational singularities \cite{labs_10}\\
      challenge\_12b    &    40 & many candidates to check \cite{labs_10}\\
      mignotte\_xy      &    42 & a product of $x$/$y$-Mignotte polynomials, displaced;\newline many clustered solutions\\


      \hline
    \end{tabularx}
  \end{minipage}
\end{table}